\documentclass[sigconf]{acmart}

\usepackage{amssymb}
\usepackage{stmaryrd}
\usepackage{amsmath}
\usepackage{graphicx}
\usepackage{subfigure}
\usepackage{caption}
\usepackage[ruled,lined]{algorithm2e}
\usepackage{tikz}
\usepackage{mathtools}
\usepackage{multirow}

\newcommand{\share}[1]{\llbracket{#1}\rrbracket}

\newcommand{\Ind}{\mathrm{Ind}}

\title{Secure Machine Learning over Relational Data}

\author{Qiyao Luo$^1$ $\quad$ Yilei Wang$^1$ $\quad$ Zhenghang Ren$^1$ $\quad$Ke Yi$^1$ $\quad$Kai Chen$^1$ $\quad$Xiao Wang$^2$}
\affiliation{%
  \institution{$^1$The Hong Kong University of Science and Technology \\
  $^2$Northwestern University}
}
\email{(qluoak, ywanggq, zrenak, yike, kaichen)@cse.ust.hk; wangxiao@cs.northwestern.edu}

\begin{abstract}
    A closer integration of machine learning and relational databases has gained steam in recent years due to the fact that the training data to many ML tasks is the results of a relational query (most often, a join-select query).  In a federated setting, this poses an additional challenge, that the tables are held by different parties as their private data, and the parties would like to train the model without having to use a trusted third party.  Existing work has only considered the case where the training data is stored in a flat table that has been vertically partitioned, which corresponds to a simple PK-PK join.  In this paper, we describe secure protocols to compute the join results of multiple tables conforming to a general foreign-key acyclic schema, and how to feed the results in secret-shared form to a secure ML toolbox.  Furthermore, existing secure ML systems reveal the PKs in the join results.  We strengthen the privacy protection to higher levels and achieve zero information leakage beyond the trained model.  If the model itself is considered sensitive, we show how differential privacy can be incorporated into our framework to also prevent the model from breaching individuals' privacy. 
\end{abstract}

\begin{document}

\maketitle

\newcommand{\PK}{\mathrm{PK}}

\section{Introduction}
\label{sec:intro}
A common task is in privacy-preserving machine learning is \textit{learning over vertically partitioned data} \cite{qiang19flca,ppregression,ppvfl}.  Here, the training data is given as a relational table, where one column is the primary key (PK), one column is the label (for supervised learning), while the rest of the columns are the features.  Data in this table is contributed by two parties, Alice and Bob.  More precisely, the PK column is shared, but each of the other columns is owned by one of the two parties. Yet, collectively, they would like to train a model without having to reveal their own columns to the other party.  Note that while both Alice and Bob have the PK column, they do not necessarily have the same set of PKs.  This is common in practice, because a party might be only willing to contribute a subset of her data for the purpose of training a particular model, and different parties may have different rules (i.e., predicates) in selecting which records to contribute.  In this case, the training shall be done on the common records (i.e., the records with matching PKs) of the parties' data, or the results of a \textit{join-select} query, if using database terminology. 

\begin{table}
\small
\begin{tabular}{@{\hskip 0pt}c|cccccc@{\hskip 0pt}}
\hline
Level & Accuracy &Model & Join size & Join PK & Gradient & Ref\\
\hline
1 & High & Yes & Yes & Yes & Yes & \cite{qiang19flca,hardy17}\\
2 & High & Yes & Yes & Yes & No & \cite{secureml,helen}\\
3 & High & Yes & Yes & No & No & this work\\
4 & High & Yes & No & No & No & this work\\
5 & Med/Low & DP & No & No & No & this work\\
\hline
\end{tabular}
\caption{Different levels of privacy protection in terms of information leakage.}
\label{tab:compare}
\vspace{-2em}
\end{table}

Depending on what information is revealed, different techniques have been proposed in the literature, as summarized in Table~\ref{tab:compare}, where higher levels correspond to stronger privacy protection.
Level 1 protocols are a common practice, which hide the original data, but reveal the gradients.  This offers minimal privacy protection, since it is easy to derive the original data from the gradients, especially when there is some background knowledge, such as the fact that the labels must be $0$ or $1$.  

Using \textit{secure multi-party computation (MPC)}, SecureML \cite{secureml} and Helen \cite{helen} achieve level 2 privacy protection, where the gradients are hidden from either party.  Both level 1 and level 2 protocols first perform encrypted entity alignment \cite{ppschemadm,ppintdbop} to identify the common records for training, which reveal the PKs in the join results to both parties (PKs not in the join results are not revealed).  Although the PKs  themselves may not carry much information (they are a subset of the PKs already held by the two parties anyway), their \textit{presence} or \textit{absence} in the join results can reveal sensitive private information.  

\begin{example}
\label{ex:1}
Consider a scenario where an insurance company and a hospital would like to collectively train a model, say, to predict customers' preferences of insurance policies based on their medical history.  The PKs in this case could be the customers' ID numbers or passport numbers.  Then knowing that a particular customer is in the hospital's database is already breaching the customer's privacy.  The situation gets worse if the parties would like to train a model using a subset of their data satisfying a particular predicate (the predicate itself is known to both parties).  Suppose the two parties decide to  train a model on patients with a particular type of disease, or on customers having insurance policies with a premium $\ge \$1,000$.  In the former case, the insurance company would know which of their customers have this disease; in the latter case, the hospital would know which of their patients do \textit{not} have enough insurance coverage.
\end{example}

\paragraph{Stronger privacy protection}
In this work, we aim at higher levels of privacy protection in terms of information leakage.  We design techniques that not only hide the PKs in the join results (level 3), but also the join size (level 4).  
Note that the two address privacy concerns at different levels: Revealing the PKs in join result breaches the privacy of individuals, while revealing the join size breaches the privacy on the party level, especially when predicates are used.  Continuing with Example \ref{ex:1}, revealing the join size will let the hospital know the number of patients with high-premium insurance policies, which could be sensitive information that the insurance company is not willing to reveal.  From a theoretical point of view, only level 4 (or higher levels) meets the rigorous security definition of the MPC model, i.e., zero information is leaked other than the output of the functionality, which is the trained model in this case.

In levels 1--4, the trained model is accurate, i.e., it is the same as the one that would have been obtained if all data were shipped to a trusted third party for training (except for some negligible losses due to calculating the gradients with finite-precision arithmetic in cryptographic operations \cite{secureml}).  As demonstrated in \cite{zhu2019deep,song2017machine}, an accurate model may be used to extract sensitive personal information (although this does not violate the security definition of the MPC model), so we also provide the option of using \textit{differential privacy (DP)} \cite{Abadi_2016, NEURIPS2018_7221e5c8} to inject noise to the model, achieving the highest level of privacy protection.   However, this level of protection inevitably incurs certain losses to the model accuracy, depending on the amount of noise injected, which in turn depends on the privacy requirement (i.e., the DP parameters).  Note that level 5 hits the fundamental trade-off between privacy and utility.  On the other hand, for levels 1--4, the trade-off is between the training costs (time and communication) and the privacy.

\paragraph{The relational model}
We also significantly broaden the data model.  Prior work has only studied the simplest case where the training data is defined as the result of a PK-PK join, which is just a set intersection problem. In this work, we adopt a much more general, relational model.  More precisely, we consider a model where the training data is stored in separate tables according to a \textit{foreign-key acyclic schema} (precise definition given in Section~\ref{sec:schema}).  The tables are owned by different parties, who collectively would like to train a model over the join results of all the tables, possibly after applying some predicates each party may impose on their tables (type-1 predicates), or even some predicates that span multiple relations from different parties that cannot be evaluated by either party on their own (type-2 predicates).

The relational model easily incorporates the PK-PK join as a special case, while offering generality that can be useful in a variety of federated learning scenarios.

\begin{example}
\label{ex:bank}
Suppose a bank stores its customers' data as a table $\mathtt{Account}(\mathtt{personID}$, $\mathtt{age}$, $\mathtt{balance}$,  $\dots)$ and an e-commerce company has the transaction records in a table 
$\mathtt{Transactions}$ $(\mathtt{sellerID, buyerID, product, price})$. We assume that the two companies use consistent IDs as people's identifiers (otherwise, joint training is impossible).  Although there are only two physical tables, the join results are actually defined by a three-table join: 
\begin{align*}
&\mathtt{Account1}(\mathtt{sellerID},\mathtt{age1}, \mathtt{balance1})  \Join\\
&\mathtt{Transactions}(\mathtt{sellerID, buyerID, product, price})\Join \\
 &\mathtt{Account2}(\mathtt{buyerID},\mathtt{age2}, \mathtt{balance2}),
\end{align*}
where $\Join$ denotes natural join.  Note that the join involves two logical copies of $\mathtt{Account}$ with different attribute renamings.  From the join results, many interesting models can be trained, e.g., to predict the likely products a buyer tends to buy from a seller, given the buyer's balance and the seller's age.  The bank may impose a type-1 predicate $\mathtt{balance2 > \$100,000}$ so that the model only considers such high-value customers.  An example of a type-2 predicate is $\mathtt{balance1 > balance2}$.  Note that although the bank has both attributes (they are actually the same physical attribute), it cannot evaluate the predicate because it does not know if any two customers will appear together in the join results, which is known only to the e-commerce company.  The naive way of precomputing the Cartesian product of the table $\mathtt{Account}$ and filtering the Cartesian product with the predicate would increase the table size quadratically.
\qed
\end{example}

The previous example only involves two parties and three logical tables (thereafter ``tables'' always mean ``logical tables''), while more complex scenarios can certainly be imagined with more parties and tables.  For simplicity, in this paper we primarily focus on the case with two parties but any number of tables; extension to more than two parties is discussed in Section~\ref{sec:twoserver}.

Note that under the relational model, it is more important to hide the PKs in the join results.  For a  simple PK-PK join, the appearance of a PK in the join results is binary.  For a general join, the multiplicity of a PK in the join results may convey more information.  In Example~\ref{ex:bank}, revealing the PKs in the join results will let the bank know how many transactions each customer has made, either as a seller or as a buyer.
The join size is less sensitive, but it still contains information sensitive about a  party, especially when predicates are used.

\medskip
To summarize, we make the following contributions in this work:
\begin{enumerate}
\item We formulate the problem of secure machine learning over relational data, bridging an important gap between privacy-preserving machine learning and relational databases.
\item By employing MPC techniques, we design  protocols against semi-honest adversary to compute the join results of a relational database conforming to a foreign-key acyclic schema, where the tables are arbitrarily partitioned to two parties. Both the running time and communication cost are near-linear in the data size. 
Existing techniques can only deal with PK-PK joins; for PK-FK joins, they either reveal extra information (e.g., key frequencies \cite{fdj}) or require the key frequencies to be bounded by a constant \cite{senate}, while our method does not reveal any information or require any constraints.
\item In addition to join, we also show how to support other relational operators, including selection, projection, and group-by aggregation, to further process the join results before feeding them (in secret-shared form) to a  secure machine learning toolboxes such as SecureML \cite{secureml}.  
The whole process provides level 3 or 4 privacy protection.
\item We achieve level 5 privacy protection by combining with techniques from differential privacy, but at the (inevitable) loss of some accuracy of the trained model.
\item We have built a system prototype and demonstrated its practical performance with a variety of relational schemas and machine learning tasks.
\end{enumerate}

\section{Related Work}
The MPC model was first conceptualized by Yao in his pioneering paper \cite{yao1982protocols}.  Protocols in this model generally fall into two categories: generic and customized. Generic protocols, such as \textit{Yao's garbled circuits} \cite{yao1986generate}, \textit{GMW} \cite{goldreich1987how}, and \textit{BGW} \cite{benor1988completeness}, are based on expressing the computation as an arithmetic or Boolean circuit.  For certain problems, such as sorting and compaction (see Section \ref{sec:sort}), a circuit-based protocol is still the best choice, both in theory and in practice.  Fairplay \cite{fairplay} provides a programming language and a compiler, which can compile a user program into a circuit that can be securely executed.   However, for some other operations, such as set intersection (see \ref{sec:psi}), circuit-based protocols have large costs, and customized protocols have been developed that can achieve much higher efficiency, in terms of both computation and communication.  In our system, we use a mix of generic, circuit-based protocols and customized protocols to achieve the best overall performance. 

SMCQL \cite{bater2017smcql} is the first query processing engine in the two-party MPC model, assuming semi-honest parties. The engine contains a query planner and a secure executor.  The query planner generates a garbled circuit based on the query, which is then executed by the secure executor.  It does not make use of any customized MPC protocols.  In particular, they implement a multi-way join using a naive circuit of size $O(N^k)$, where $k$ is the number of tables and $N$ is the size of each table. Thus SMCQL cannot scale to joins involving more than a few hundred tuples.  Senate \cite{senate} uses customized protocols for PK-PK joins, thus achieving a significant improvement.  However, when generalizing to PK-FK joins, which is the focus of this paper, their protocol needs an upper bound $\tau$ on the multiplicities of the FKs. In Example \ref{ex:bank}, this corresponds to the maximum number of transactions a customer can be involved as a buyer and as a seller.  Its performance quickly deteriorates as $\tau$ increases.  In the worst case, their join protocol degenerates to the naive garbled circuit.   On the other hand, our method does not need such an upper bound while having a near-linear complexity irrespective of the FK multiplicities.   Secure Yannakakis \cite{secyan} provides a secure version of the classical Yannakakis algorithm for computing $\alpha$-acyclic join-aggregate queries. Its computation and communication costs are linear in data size.  However, their protocol only supports join-aggregate queries, but not machine learning.  Also, FK-acyclicity (which we adopt) and $\alpha$-acyclicity are different notions.  For example, the TPC-H schema is FK-acyclic but not $\alpha$-acyclic.  For machine learning tasks, the former is more widely used, as most data warehouses adopt a star or a snowflake schema, which are special cases of FK-acyclic schemas. 
\citet{fdj} provide a protocol for database joins on secret shared data in the honest-majority three-party setting. Their protocol only supports PK-PK joins; for PK-FK joins, the data distribution of some tables will be revealed. 

The goal of federated learning \cite{qiang19flca,fedopt,fedlearninggoogle,li2020federated} is to train a model from distributed and private data.  MPC and differential privacy provide the most rigorous definitions of privacy in this setting.  
SecureML \cite{secureml}, which we review in detail in Section \ref{sec:secureml}, provides the first two-party  MPC protocols against semi-honest adversary for linear regression, logistic regression, and neural networks, while achieving level 2 privacy protection.  Helen \cite{helen} trains linear models over more than two parties and can defend against malicious adversaries, 
while also achieving level 2 privacy protection. Our system is built on top of SecureML and achieves level 3--5 in terms of information leakage.  In terms of the power of the adversary, it is weaker than that of \cite{helen}, but in principle, our protocols can also be hardened to defend against malicious adversaries.
Like our level 5 privacy protection, \citet{NEURIPS2018_7221e5c8} inject DP noise to the trained model on top of MPC.  However, they consider the case where the training data is stored in a flat table horizontally partitioned over multiple parties, i.e., no joins.

Learning over joins has received much attention recently \cite{factorized,mloverrd,mloverrdbftu,ac/dc}.  The motivation of this line of work is similar to ours, i.e., training data is often stored in separate tables connected by PK-FK references.  They try to improve over the naive approach, which computes the join first and then feed the join results to an ML system.  In our case, this naive approach does not work at all, as the tables are held by different parties as their private data, which poses a completely different set of challenges.






\section{Preliminaries}

\subsection{Foreign-key Acyclic Schema}
\label{sec:schema}
Let $\mathbf{R} = ( R_0(\mathbf{x}_0), R_1(\mathbf{x}_1), \dots, R_k(\mathbf{x}_k))$ be a relational database.  Each $R_i(\mathbf{x}_i)$ is a table with attribute set $\mathbf{x}_i=\{x_{i1}, x_{i2},\dots\}$. Let $\PK(R_i)$ be the primary key of $R_i$; we also use the notation $R_i(\underline{x_{i1}}, x_{i2}, \cdots)$ to indicate  $\PK(R_i)=x_{i1}$. For a tuple $t\in R$, we use $\PK(t)$ to denote the value of $t$ on $\PK(R)$. All tuples in $R$ must have unique values on $\PK(R)$. For a composite PK, we (conceptually) create a combined PK by concatenating the attributes in the composite key, while the original attributes in the composite key are treated as regular, non-key attributes.  For example, the table ${\tt partsupp}$ in the TPC-H schema has a composite PK ${\tt (\underline{ps\_partkey}, \underline{ps\_suppkey})}$. We rewrite it as ${\tt partsupp(\underline{ps\_partsuppkey},ps\_partkey, ps\_suppkey}$, $\dots)$.

An attribute $x_{ik}$ of a table $R_i$ is a {\em foreign key (FK)} referencing the PK $\underline{x_{j1}}$ of table $R_j$, if the $x_{ik}$ values taken by tuples in $R_i$ must appear in $\underline{x_{j1}}$. If a table has a composite FK, we similarly create a combined attribute referencing the composite PK. For example in the TPC-H schema, the table {\tt lineitem} has a composite FK {\tt(l\_partkey,l\_suppkey)}.  We create a new attribute {\tt l\_partsuppkey} referencing {\tt ps\_partsuppkey}.  

\begin{figure}[h]
\centering
\begin{tikzpicture}

  \node[anchor=west] (p1) at ( 0, 0) {lineitem}; 
  \node[anchor=west] (p2) at ( 1.7, -0.4) {orders};
  \node[anchor=west] (p3) at ( 1.7, 0.4) {partsupp};
  \node[anchor=west] (p4) at ( 3.4, 1) {part};
  \node[anchor=west] (p5) at ( 3.4, 0.4) {supplier};
  \node[anchor=west] (p6) at ( 3.4, -0.4) {customer};
  \node[anchor=west] (p7) at (5.1, 0) {nation};
  \node[anchor=west] (p8) at (6.4, 0) {region};
  
  \begin{scope}[every path/.style={<-}]
    \draw (p2) -- (p1);
    \draw (p3) -- (p1); 
    \draw (p4) -- (p3);
    \draw (p5) -- (p3);
    \draw (p6) -- (p2);
    \draw (p7) -- (p6);
    \draw (p7) -- (p5);
    \draw (p8) -- (p7);
  \end{scope} 
\end{tikzpicture}
\caption{The foreign-key graph of TPC-H schema.}
\label{fig:TPCHSchema}
\end{figure}
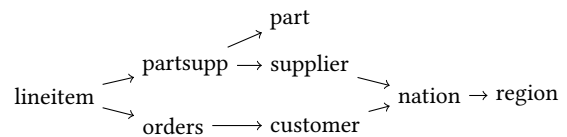

We can model all the PK-FK relationships as a directed graph: Each vertex represents a table, and there is a directed edge from $R_i$ to $R_j$ if $R_i$ has an FK referencing the PK of $R_j$.  We call this graph the \textit{foreign-key graph}.  
A basic principle in relational schema design is that this graph should be acyclic; one example is the TPC-H schema (Figure~\ref{fig:TPCHSchema}).  Cycles in the PK-FK relationships would create various integrity and quality issues in data maintenance, and should always be avoided.  As in most data warehouses, we assume that there is exactly one table with in-degree $0$ (i.e., no other tables reference this table via FKs), which is often called the \textit{fact table}, while the other tables the \textit{dimension tables}.  We use $R_0$ to denote the fact table, while $R_i, i\in [n]=\{1,\dots, n\}$ are the dimension tables.  For example, $\mathtt{Transactions}$ is the fact table in Example~\ref{ex:1}, while the fact table in the TPC-H schema is $\mathtt{lineitem}$.  Note that a snowflake schema is a special acyclic schema where the foreign-key graph is a tree.  For two tables $R_i$ to $R_j$, if there is directed path in the foreign-key graph from $R_i$ to $R_j$, we say that $R_i$ is an ancestor of $R_j$. Similarly, if a tuple $t_i \in R_i$ can join with $t_j \in R_j$ through the FK references directly or indirectly, we say that $t_i$ is an ancestor of $t_j$. In particular, we consider a tuple to be an ancestor of itself.

We assume that the training data is the join results of all the tables, possible with some predicates.  The join conditions are all the PK-FK pairs, i.e., for each FK $x_{ik}$ in $R_i$ referencing the PK $x_{j1}$ in $R_j$, we have a join condition $x_{ik} = x_{j1}$.  An important departure we make from the strict relational model is that FK constraints are not enforced in our data model.  These constraints are meant to ensure the integrity of data inside one organization, but they do not make much sense when different tables are owned by different parties.  For instance, in the schema of Example~\ref{ex:1}, it may not be the case that the sellers and buyers must exist in the bank's database.  Essentially, we use the PK-FK relationships only to define the join conditions, but do not enforce the FK constraints.  On the other hand, we do require that the PKs must be unique, which still makes sense in the multi-party setting since this is a constraint on individual tables.  This implies that each tuple in the fact table appears at most once in the join results, while a tuple in a dimension table may appear zero, one, or any number of times.

In addition, the parties may use a set of predicates to select a subset of the join results for the purpose of training a particular model.  There can be two types of predicates. (1) If a predicate only involves attributes from one table, then the party that owns the table can perform the filtering as a preprocessing step.  (2) If a predicate involves attributes from different tables, we defer its processing after the join.  

We use $J$ to denote the join results (after applying the predicates  if there are).  For an FK acyclic schema, we always have $|J| \le |R_0|$.

\subsection{Secure Multi-Party Computation (MPC)}
Secure Multi-Party Computation (MPC) protocols allow parties to jointly compute a function without revealing their own information.  We focus on the setting with two parties $P_0$ and $P_1$, also often named as Alice and Bob. Suppose $P_i$ has input $x_i$, $i=0,1$.  In MPC, the goal is to design a protocol for Alice and Bob to compute $(y_0,y_1) \leftarrow F(x_0, x_1)$ for some given functionality $F$, so that at the end of the protocol, $P_i$ learns nothing more than $y_i$. A more rigorous security definition using the real-ideal world paradigm can be found in \cite{SMCBOOK}; the formal definition is not relevant to our development, as we will not invent new security protocols, but only use existing ones with proven security guarantees.  In particular, we will adopt the \textit{semi-honest} model, where the parties will faithfully execute our protocol, but may try to learn as much information as possible from the transcript.  Our algorithms can also be hardened to defend against malicious adversaries (i.e., the parties may deviate from the protocol), by replacing the corresponding security primitives with their malicious counterparts, but the costs will also increase significantly. 

During an MPC computation, intermediate and final results are often stored in a \textit{secret-shared} form.
In the two-party setting, a value $v \in \mathbb{Z}_n$ can be shared as $v = (\share{v}_0 +\share{v}_1) \bmod n$, where each $\share{v}_i, i\in \{0,1\}$ is a uniform (but not independent, obviously) number in $\mathbb{Z}_n$. It is clear that each share by itself reveals nothing about $v$, but they reconstruct $v$ when combined together. It is also easy to convert a value to its shared form and vice versa.
Addition and multiplication can be performed in shared forms without reconstructing the plaintext.  Given two values in shared form $\share{x}$ and $\share{y}$, it is easy to verify that $(\share{x}_0 + \share{x}_1) + (\share{y}_0 + \share{y}_1) \equiv x+y \pmod{n}$, i.e., $(\share{x}_0 + \share{x}_1) \bmod n$ and $(\share{y}_0 + \share{y}_1)\bmod n$ form a valid sharing of $z=x+y$, so addition over shared values can be done easily without communication. Multiplication is more complicated and requires communication and some cryptographic operations.  \textit{Multiplication triples} \cite{beavertriples} are a common technique to mitigate the high cost of multiplications.  It pushes most of the cost to a data-independent, offline stage.  During an MPC computation, one can compute a sharing of $x\cdot y$ from $\share{x}$ and $\share{y}$ by consuming one precomputed multiplication triple.

The above sharing scheme, also known as \textit{arithmetic sharing}, works well with an arithmetic circuit (i.e., each gate is either an addition or a multiplication over $\mathbb{Z}_n$).  This is inconvenient for Boolean operations, such as comparisons.  Furthermore, as each multiplication requires a round-trip message, the number of communication rounds would be proportional to the depth of the circuit, which may be an issue for high-latency communication networks, such as over the Internet. To address these issues, two other sharing schemes, known as \textit{Boolean sharing} and \textit{Yao's sharing}, can be used instead.  The former works well with a Boolean circuit, while the latter requires a constant number of communication rounds regardless of the circuit depth (a.k.a.~\textit{Yao's garbled circuits} \cite{yao1982protocols}).  Furthermore, there are techniques to convert between these sharing schemes \cite{demmler2015aby}.  The details of these sharing schemes and how the conversion works are not needed to understand the rest of the paper.  When describing our algorithms, we will also not explicitly state which sharing scheme is used, with the understanding that the most appropriate one will be used.

While secret-sharing protects private data and intermediate results, the access pattern of an algorithm may also leak sensitive information, and should be made \textit{oblivious} to the input data.  One way to design an oblivious algorithm is to express the computation as a circuit, which is clearly data-independent. However, for a multi-way join, a naive circuit would perform a comparison for each of the $\prod_{i=1}^k |R_i|$ combinations of tuples, one from each table.  Such an algorithm clearly does not work beyond a few tiny tables.

In this paper, when using the $O(\cdot)$ notation to express an algorithm's running time and communication cost (henceforth the term ``cost'' includes both running time and communication cost), we treat each arithmetic or Boolean operation over shared values, as well as a conversion between different sharing schemes, as an atomic operation.  In practice, these costs depend on the security parameters and the bit length of the integers being manipulated.  


\subsection{Secure Machine Learning}
\label{sec:secureml}
Recall that in MPC, the goal is to compute $(y_0,y_1)\gets F(x_0,x_1)$. In secure machine learning, the inputs $x_0$ and $x_1$ are the training data (in our case, $x_0, x_1$ are a partitioning of the relational tables), $F$ is the training algorithm, and $y_0=y_1$ is the trained model.  In this paper, we consider training algorithms that follow the popular minibatch stochastic gradient descent (SGD) framework. 

More precisely, the model is defined by a coefficient vector $\mathbf{w}\in \mathbb{R}^d$ where $d$ is the number of features.  Given a loss function $L$, minibatch SGD starts from a randomly initialized $\mathbf{w}_0$, and iteratively updates it as 
$$\mathbf{w}_{i+1} \gets \mathbf{w}_i - \frac{\eta}{|B|} \sum_{t \in B}  \nabla L(t, \mathbf{w}_i),$$
where $B$ is a random sample of training data ($|B|$ is called the batch size) and $\eta$ is the learning rate. 

SecureML \cite{secureml} provides several minibatch SGD instantiations, including linear regression, logistic regression, and neural networks, under the MPC model. The inputs $x_0$ and $x_1$ are two tables with a common PK column.  Before the training starts, SecureML reveals the common PKs so as to link the features with the labels.  Then, it converts $\mathbf{w}_0$, the features, and labels into shared form and runs minibatch SGD to calculate $\mathbf{w}_i, i=1,\dots, T$, all in shared form.  The main technical contribution of SecureML is to show how to perform floating-point arithmetic and how to compute exponentiation (as needed in computing the gradient for logistic regression) in shared form with little loss in accuracy. It also invents a technique to save multiplication triples when the same value is involved in many multiplication operations, as in the case when computing $\sum_{t \in B}  \nabla L(t, \mathbf{w}_i)$.  The obliviousness of the algorithm, on the other hand, is trivially achievable, since each training data record in $B$ goes through exactly the same sequence of operations. 
In non-private machine learning, SGD would iterate until convergence, but for the MPC model, the number of iterations may also be sensitive information.  To address the issue, SecureML fixes the number of iterations $T$ in advance, and we will adopt the same strategy.

SecureML will compute the trained model in shared form.  Then, depending on the agreement between the two parties, the model can be revealed to either or both parties, or remain in secret-shared form forever.  In the latter case, no party ever obtains the model, but they can still jointly evaluate the model inside MPC on any testing data whenever required \cite{secureml}.

\subsection{Differential Privacy}
While MPC ensures that the transcript of the protocol does not leak any private information, the output of the functionality $F$ is still revealed, which may contain private information.  To address this issue, a popular framework is \textit{differential privacy (DP)} \cite{dwork2014algorithmic}.  We adopt the DP policy defined in \cite{kotsogiannis2019privatesql} for relational data following an FK acyclic schema.  Given such a schema $\mathbf{R}$, one of the tables, say $R_p$, is designated as the \textit{primary private table}, while any other ancestor of $R_p$ in the FK graph is called a \textit{secondary private table}.  For instance, one may designate $\mathtt{customer}$ as the primary private table, then $\mathtt{orders}$ and $\mathtt{lineitem}$ will be secondary private tables. Two database instances $\mathbf{I}$ and $\mathbf{I}'$ are said to be \textit{neighboring instances}, denoted $\mathbf{I} \sim \mathbf{I}'$, if one can be obtained from the other by deleting a set of tuples, all of which are ancestors of the same tuple $t\in R_p$.  Then an algorithm $M$ is $(\varepsilon, \delta)$-differential private if for any $\mathbf{I}\sim\mathbf{I}'$ and any set of outputs $Y$, 
$$\Pr[M(\mathbf{I}) \in Y] \leq e^{\varepsilon} \cdot \Pr[M(\mathbf{I}') \in Y] + \delta.$$
Here $\varepsilon$ and $\delta$ are the privacy parameters, where smaller values correspond to stronger privacy guarantees.  In practice, $\varepsilon$ often takes a value between $0.1$ and $10$, while $\delta$ should be smaller than $1/N$, where $N$ is the total number of tuples of all the tables.  The intuition behind this definition is that the trained model would be statistically indistinguishable if part or all of the data associated with any one tuple in the primary private table had been deleted from the database.  For example, when $\mathtt{customer}$ is the primary private table, this implies that one is unable to tell if any particular customer had withdrawn part or all of her associated orders and lineitems from the database. 

Note that the privacy protection of DP is (inevitably) weaker than that of MPC: The former is on the level of individuals, and even so, the protection is only to a level controlled by $\varepsilon,\delta$; while the latter protects all data owned by all parties, which includes data of many individuals, and the protection is absolute (only subject to standard cryptographic assumptions).

\section{Cryptographic Primitives}

In this section, we describe some cryptographic primitives, as well as how to adapt them to suit the need of our protocols.

\subsection{Oblivious Extended Permutation (OEP)}
\label{sec:oep}
Suppose Alice holds a function $\xi:[N]\rightarrow[M]$, and Bob holds a length-$M$ sequence $\{x_i\}_{i=1}^M$ where each $x_i \in \mathbb{Z}_n$. In the \textit{oblivious extended permutation (OEP)} problem, Alice and Bob wish to permute the sequence $\{x_i\}_{i=1}^M$ into a length-$N$ sequence $\{y_i\}_{i=1}^N$ as specified by $\xi$, i.e., $y_i=x_{\xi(i)}$. We call $\xi$ an extended permutation function. In order to keep both $\{x_i\}$ and $\xi$ private, the output $\{y_i\}$ must be obtained in a shared form. The OEP protocol of \citet{mohassel2013hide} solves the problem with $O(M\log M+N\log N)$ cost, which can be generalized to the case where $\{\share{x_i}\}$ is given in shared form while $\xi$ is still held by Alice in plaintext \cite{secyan}.

A \textit{random shuffle} \cite{10.1007/978-3-030-64840-4_12} of a sequence $\{x_i\}_{i=1}^M$ can be implemented by two OEPs: the sequence is first permuted by a random permutation function $\xi_0 : [M]\rightarrow [M]$ generalized by Alice in private, and then by another random permutation function $\xi_1$ generated by Bob in private.  Thus, the sequence is eventually permuted according to $\xi_1 \circ \xi_0$, but neither party knows anything about this composed permutation.  

\subsection{Private Set Intersection (PSI)}
\label{sec:psi}
The \textit{private set intersection (PSI)} problem is a classical MPC problem.
Suppose Alice has a set $X$ with size $M$ and Bob has a set $Y$ with size $N$; the set cardinalities $M$ and $N$ are public but the elements in the sets are not.  In the PSI problem, they wish to compute $X\cap Y$.  Earlier PSI protocols often reveal the elements in the intersection, which are the output of the PSI functionality by definition.  In our setting, however, these elements are intermediate results and must be protected.  Thus, we choose the recent protocol of \citet{pinkas2019efficient}, which has cost $O(M+N)$ and returns $M$ shared indicators $\{\share{\Ind(x \in Y)}\mid x\in X\}$.  The protocol also supports \textit{payload sharing}: Suppose Bob has a payload $W(y)$ for each $y\in Y$. In addition to the indicators, the protocol also outputs $M$ shared payloads $\{\share{W(x)}\mid x\in X\}$, where $W(x)$ is a random number when $x\notin Y$. 
It can also be generalized to the case where the payloads are given in shared form, although $X$ and $Y$ must still be held by Alice and Bob in plaintext respectively \cite{secyan}, in which case the cost becomes $O(M\log M+N\log N)$.

\subsection{Sorting and Compaction}
\label{sec:sort}
Sorting is another classical MPC primitive.  Here we are given a sequence $\{\share{x_i}\}_{i=1}^N$  in shared form, and the goal is to obtain $\{\share{y_i}\}_{i=1}^N$, such that $\{y_i\}_{i=1}^N$ is a permutation of $\{x_i\}_{i=1}^N$ and $y_i \leq y_{i+1}$ for $i=1,2,\cdots,n-1$. Theoretically optimal protocols based on the AKS sorting network \cite{AKSSorting} achieve $O(N \log N)$ cost but with a huge hidden constant.  In this paper we use the one based on the bitonic sorting network \cite{bitonicsort}, which has cost $O(N \log^2 N)$ but is more practical. 

A special case of the sorting problem is to sort by $N$ binary values in shared form, namely, we simply want to place the $0$'s in front of the $1$'s.  This problem is known as \textit{oblivious compaction}, which has a simple $O(N\log N)$-cost solution \cite{compactioncircuit}.

\section{Join Protocols}
We divide our protocol into two stages.  In this section, we present our protocol to compute the join results in shared form.  In Section~\ref{sec:ML}, we show how to feed the join results into SecureML while achieving level 3 or 4 privacy protection, as well as how to use differential privacy to achieve level 5 privacy protection.  

In computing the join, we assume that the schema is public knowledge, which includes table names, attribute definitions, PK-FK constraints, and table sizes.  If a party does not wish to publicize the size of a table (say, after imposing a sensitive predicate to select a subset of the tuples from the table for the training), s/he can pad \textit{dummy} tuples to a certain size that s/he deems as safe, while the original tuples are called \textit{real}.  For each table (input, intermediate, or final join results), we add a special column $T$ to indicate whether a tuple is dummy ($T=0$) or real ($T=1$).  These indicators are stored in plaintext for an input table, but in shared form for each intermediate table or the final join results.  For a table $R$, we use $|R|$ to denote its size (possibly after padding some dummy tuples).  Recall that $R_0$ is the fact table, while $R_i, i\in [k]$ are the dimension tables.

When we say that a party, say Alice, holds a table $R(A,\share{B})$, this means that she has the values of attribute $A$ in plaintext, while the values in $B$ are in shared form.  More precisely, Alice holds two sequences $\{ a_i\}_{i=1}^{|R|}$ and $\{ \share{b_i}_0\}_{i=1}^{|R|}$, while Bob holds  $\{ \share{b_i}_1\}_{i=1}^{|R|}$.  Note that the orders of the two sequences must match.  We also generalize the notation to multiple attributes as $R(A_1, \dots, A_{k_A},\share{B_1}, \dots, \share{B_{k_B}})$ in the natural way.

\subsection{Binary Join}
\label{sec:binary}
A basic building block of our protocol is one that computes a binary join $J(A,B,C)=R(\underline{A}, B) \Join S(\underline{B}, C)$, where $R.B$ is an FK referencing $S.B$.  Here $\Join$ denotes natural join, i.e., the implicit join condition is $R.B=S.B$.  If $R$ and $S$ are held by the same party, the join can be computed in plaintext.  Below, we assume Alice holds $R$ while Bob holds $S$.  

Let $M=|R|, N=|S|$, and let $R=\{(a_i, b_i^R)\}_{i=1}^M$, $S=\{(b_i^S, c_i)\}_{i=1}^N$.  Since $B$ is the primary key of $S$, each tuple of $R$ can join with at most one tuple in $S$, so $|J| \le M$. Since the size of $J$ is sensitive to the input data, we will output a table of size exactly $M$, by padding dummy tuples if necessary. Specifically, our protocol will output a table  $J(A,B,\share{C},\share{T}) = \{(a_i,b_i^R,\share{c_i^R},\share{t_i})\}_{i=1}^M$ to Alice.  For each $i\in [M]$, if $b_i^R = b_j^S$ for some $j\in [N]$, then $c_i^R=c_j$ and $t_i = 1$; otherwise $c^R_i$ is a random number and $t_i=0$.  Note that for Alice, the table $J(A,B,\share{C},\share{T})$ contains no more information than her input table $R(A,B)$, while Bob only receives $\{(\share{c_i^R}_1, \share{t_i}_1)\}_{i=1}^M$, which contains no information, either.

Our protocol runs in a constant number of rounds with $O(M\log M + N\log N)$ cost.  It proceeds in two steps, as follows.

\begin{enumerate}
    \item \textbf{Set intersection:} Alice and Bob first run PSI with payload.  Alice inputs a sequence $X$ that consists of all the distinct values in $R.B$.  Since $|\pi_B(R)|$, the number of distinct values in $R.B$, is sensitive information, Alice pads dummy elements so that $X$ consists of exactly $M$ elements.  More precisely, we have $X=\{x_i\}_{i=1}^M$, where $x_i$ is the $i$-th distinct value in $R.B$ for $1\le i \le |\pi_B(R)|$, and $x_i$ is a dummy element for $i> |\pi_B(R)|$.  Bob inputs a sequence $Y=\{b_i^S\}_{i=1}^N$ where each $b_i^S$ has payload $W(b_i^S) = c_i$.  The PSI protocol would return indicators $\{\share{\Ind(x_i)}\}_{i=1}^M$ and payloads $\{\share{W(x_i)}\}_{i=1}^M$ in shared form, where $\Ind(x_i)=1$ if $x_i = b_j^S$ for some $j$ and $\Ind(x_i)=0$ otherwise.  In the former case, we have $W(x_i)=c_j$, while $W(x_i)$ is a random number in the latter case.
    \item \textbf{Reorder:} Let the extended permutation function $\xi: [M] \rightarrow [M]$ be such that $b_i^R = x_{\xi(i)}$.  Note that $\xi$ is sensitive information to Alice.  In the second step, we use OEP to permute $\{(\share{W(x_i)}, \share{\Ind(x_i)})\}_{i=1}^M$ according to $\xi$.   The OEP protocol would return $\{(\share{c_i^R},\share{t_i})\}_{i=1}^M$ in the same order as $R=\{(a_i, b_i^R)\}_{i=1}^M$, which can then be combined to form the desired output $J(A,B,\share{C},\share{T})$.
\end{enumerate}

The basic protocol above can be easily generalized to handle the following cases:

\paragraph{Multiple attributes}
The case where $R$ and/or $S$ have more than two attributes can be handled straightforwardly. Nothing special has to be done to any additional attributes in $R$, and we just need to put them together with the join results $J(A,B,\share{C},\share{T})$. Any additional attribute in $S$, however, has to be treated as payloads and they will appear in the join results in shared form, just like attribute $C$.

\paragraph{Shared attributes}
The protocol can also easily handle the case where $C$ is given in shared form. All we have to do is to replace the PSI protocol with the the version that works on shared payloads.  Combined with the above generalization to multiple attributes, this means that we can handle a mixture of attributes, some in plaintext while others in shared form. However, $B$ (in both $R$ and $S$) must be given in plaintext.  For example, given $R=(\underline{A}, B, \share{D})$ and $S=(\underline{B}, \share{C}, E)$, the protocol will return the join results as $J=R\Join S =(\underline{A}, B, \share{D}, \share{C},\share{E}, \share{T})$ to Alice. Note that $\share{C}$ would be re-shared with fresh randomness in $J$, while $\share{D}$ retains its original shares.

\paragraph{Existing indicators}
When $R$ and/or $S$ have been padded with dummy tuples (e.g., to hide the selectivity of a predicate) or are the  results of a previous join, they have an existing indicator attribute $T$, which may be in either plaintext or shared form. Here we only consider the most general case where both tables have an indicator and both indicators are given in shared form; the other cases are simpler. In this case, Alice has $R=(\underline{A}, B, \share{T^R})$ and Bob has $S=(\underline{B}, C, \share{T^S})$. We first treat $\share{T^R}$ and $\share{T^S}$ as an ordinary attribute of $R$ and $S$ given in shared form respectively, and compute the join $R\Join S$, which would return $J' = (\underline{A}, B, \share{C}, \share{T^R}, \share{T^S}, \share{T'})$ to Alice ($\share{T^S}$ would be re-shared with fresh randomness in $J'$), where $T'$ is the indicator of $J'$.  Then Alice and Bob compute $\share{T} =\share{T^R \cdot T^S\cdot T'}$ row by row, and produce the final join result $J=(\underline{A}, B, \share{C}, \share{T})$.

\paragraph{PK-PK join}
If $B$ is also a PK in $R$, then this becomes a PK-PK join.  In this case, we simply skip the reorder step since $\xi$ would be an identity permutation.

\paragraph{Same owner}
When $R$ and $S$ belong to the same owner (say, Alice) and there are no shared attributes, the join can be computed locally in plaintext, but we still need to pad dummy tuples and add an indicator attribute $T$, so that $J=R\Join S$ always has $|R|$ tuples.  When $R$ has shared attributes, both Alice's and Bob's shares for these attributes remain unchanged.  When $S$ has shared attributes, we need to use OEP to permute them so that they are consistent with $J$. 

\subsection{Three-way Join}
\label{threewayjoin}
We now show how to apply our binary join protocol to compute more complex joins.  We start with a three-way join.  There are two types of three-way joins: a tree join and a line join.

\begin{figure}[htbp]
    \centering
        \begin{tikzpicture}[sibling distance=5em, level distance=3em,
      every node/.style = {shape=rectangle, draw, align=center}]]
    \node [color=blue] (root) at(0, 0) {$^{P_0\ }R_1(\underline{A}, B, C)$};
    \node [color=red] (left) at(-1.5, -1) {$^{P_1\ }R_2(\underline{B}, D)$};
    \node [color=red] (right) at(1.5, -1) {$^{P_1\ }R_3(\underline{C}, E)$};
    \draw[->] (root) -- (left);
    \draw[->] (root) -- (right);
    \end{tikzpicture}
    \caption{A basic tree join.}
    \label{treebasedthreewayjoin}
\end{figure}
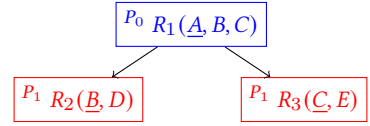

\paragraph{Tree joins}
\label{treejoin}
A basic tree join has the form $R_1(\underline{A}, B, C) \Join R_2(\underline{B}, D) \Join R_3(\underline{C}, E)$ (see Figure~\ref{treebasedthreewayjoin}).  If $R_1$ and $R_2$ (or $R_3$) are held by the same party, s/he can compute the $R_1 \Join R_2$ (or $R_1\Join R_3$) locally and the join reduces to a binary join.  Below we assume that Alice ($P_0$ in the figure) has $R_1$ while Bob ($P_1$ in the figure) has $R_2$ and $R_3$. 

We compute $R_{12} = R_1 \Join R_2 = (\underline{A}, B, C, \share{D}, \share{T_{12}})$ and $R_{13} = R_1 \Join R_3 = (\underline{A}, B, C,  \share{E}, \share{T_{23}})$ in parallel using the binary join protocol. Note that Alice has both $R_{12}$ and $R_{23}$, and they are both ordered by $A$.  Thus, Alice can simply combine the two tables, while computing $\share{T} = \share{T_{12} \cdot T_{13}}$ row by row, which produces the final  results of this three-way join as $J=(\underline{A}, B, C, \share{D}, \share{E}, \share{T})$.

If the input tables have more attributes in plaintext or shared form (the join attributes $B$ and $C$ must be in plaintext) and/or existing indicators (in plaintext or shared form), they can be handled using similar ideas as in a binary join. In the join results, attributes of $R_1$ given in plaintext will still be in plaintext, while all other attributes (including the indicator) will be shared.

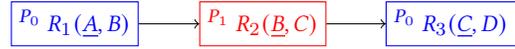
\begin{figure}[htbp]
    \centering
        \begin{tikzpicture}[sibling distance=5em, level distance=3em,
      every node/.style = {shape=rectangle, draw, align=center}]]
    \node [color=blue] (root) at(0, 0) {$^{P_0\ }R_1(\underline{A}, B)$};
    \node [color=red] (middle) at(2.5, 0) {$^{P_1\ }R_2(\underline{B}, C)$};
    \node [color=blue] (leaf) at(5, 0) {$^{P_0\ }R_3(\underline{C}, D)$};
    \draw[->] (root) -- (middle);
    \draw[->] (middle) -- (leaf);
    \end{tikzpicture}
    \caption{A basic line join.}
    \label{linebasedthreewayjoin}
\end{figure}

\paragraph{Line joins}
\label{linejoin}
A basic line join has the form $R_1(\underline{A}, B) \Join R_2(\underline{B}, C) \Join R_3(\underline{C}, D)$ (see Figure~\ref{linebasedthreewayjoin}). If $R_1$ and $R_2$ or $R_2$ and $R_3$ are held by the same party, the join reduces to a binary join, so we assume that Alice ($P_0$ in the figure) has $R_1$ and $R_3$ while Bob ($P_1$ in the figure) has $R_2$. We first use the binary join protocol to compute $R_{23} = R_2 \Join R_3 = (\underline{B}, C, \share{D}, \share{T_{23}})$.  Note that, however, the roles of Alice and Bob have flipped, and $R_{23}$ would be held by Bob, while Alice only holds her shares of $\share{D}$.  Next, we compute $J= R_1 \Join R_{23} = (\underline{A}, B, \share{C}, \share{D}, \share{T})$, and output the results to Alice.  Note that this is exactly the reason why we need our binary join protocol to be able to handle an input table with attributes and indicators in shared form.

\subsection{Foreign-key Acyclic Join}
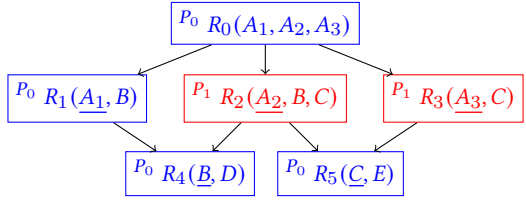
\begin{figure}
    \centering
    
    \begin{tikzpicture}[sibling distance=5em, level distance=3em,
      every node/.style = {shape=rectangle, draw, align=center}]]
    \node [color=blue] (root) at(0, 0) {$^{P_0\ }R_0(A_1, A_2, A_3)$};
    \node [color=blue] (left) at(-2.5, -1) {$^{P_0\ }R_1(\underline{A_1}, B)$};
    \node [color=red] (mid) at(0, -1) {$^{P_1\ }R_2(\underline{A_2}, B, C)$};
    \node [color=red] (right) at(2.5, -1) {$^{P_1\ }R_3(\underline{A_3}, C)$};
    \node [color=blue] (leftleaf) at(-1, -2) {$^{P_0\ }R_4(\underline{B}, D)$};
    \node [color=blue] (rightleaf) at(1, -2) {$^{P_0\ }R_5(\underline{C}, E)$};
    \draw[->] (root) -- (left);
    \draw[->] (root) -- (mid);
    \draw[->] (root) -- (right);
    \draw[->] (left) -- (leftleaf);
    \draw[->] (mid) -- (leftleaf);
    \draw[->] (mid) -- (rightleaf);
    \draw[->] (right) -- (rightleaf);
    \end{tikzpicture}
    
    \caption{Foreign-key Acyclic Join}
    \label{foreignkeyacyclicjoin}
\end{figure}

Now we are ready to describe the protocol for a general foreign-key acyclic join; see Figure~\ref{foreignkeyacyclicjoin} for such an example.  Given the DAG representing all the PK-FK relationships, our protocol iteratively executes the following steps until the DAG has only one vertex, while maintaining the invariant that all PKs and FKs in the DAG are in plaintext.  Meanwhile, we impose a set of equality constraints $\Theta$ to make sure that the join results remain the same after each step.  Initially, $\Theta=\emptyset$.

\begin{enumerate}
    \item Find a vertex with out-degree 0. This vertex must exist, otherwise the graph would be cyclic. For simplicity we assume that the table corresponding to this vertex has the form $S(\underline{B},\share{C},\share{T^S})$; the case where $S$ has more attributes can be handled similarly.  By the invariant, $B$ must be in plaintext, while the other attributes can be in either plaintext or shared form.  Let $N=|S|$, and suppose Bob holds $S$.
    \item Let $R_1,R_2,\cdots,R_d$ be the parents of $S$ in the DAG, i.e., each $R_i$ has an FK referencing $S.B$.  Again for simplicity suppose $R_i$ has the form $R_i(\underline{A_i}, B, \share{T_i})$; more attributes can be handled similarly.  There are two cases:
    \begin{enumerate}
        \item If $d=1$, we just compute $R'_1(\underline{A_1}, B, \share{C}, \share{T'_1}) = R_1 \Join S$ using the binary join protocol. The owner of $R_1$ (which may be either Alice or Bob) replaces $R_1$ with $R'_1$.  Then $S$ is removed from the DAG.
        \item If $d\ge 2$, we first replace $R_1$ with $R'_1$ as above.  Then for $i=2,3,\cdots,d$, we compute $R_i'(\underline{A_i}, B, \share{T_i'})= R_i \Join \pi_B (S)$, where $\pi_B(S)$ is computed by simply dropping the $C$ column from $S$. The owner of $R_i$ replaces $R_i$ with $R'_i$ and renames attribute $B$ into $B^{(i)}$. We then add the equality constraint $B=B^{(i)}$  to $\Theta$ and remove $S$.  Note that these equalities ensure that tuples from different $R_i$'s must join with the same tuple in $S$. 
    \end{enumerate}
\end{enumerate}

When the above process terminates with one table, we still need to check all the equalities in $\Theta$.  Note that these equalities may involve shared attributes, so we need to use a garbled circuit for each row to do the comparison.  Finally we multiply the output of this comparison circuit with the indicator in the join results to obtain the final indicator in shared form.  Note that the final join results have exactly $|R_0|$ rows, where $R_0$ is the only fact table, but some of the rows would be dummy tuples, i.e., $T=0$.

\begin{example}
\label{ex:2}
We use the schema of Figure~\ref{foreignkeyacyclicjoin} to illustrate how our protocol works, where Alice ($P_0$ in the figure) has $R_0, R_1, R_4$, and $R_5$, while Bob ($P_1$ in the figure) has $R_2$ and $R_3$.  Suppose we pick $R_4$ in the first iteration. $R_4$ has two parents $R_1$ and $R_2$.  Because Alice holds both $R_1$ and $R_4$, she can compute $R_1'(\underline{A_1}, B, D, T_1) = R_1 \Join R_4$ locally in plaintext, while adding the indicators $T_1$  and dummy tuples so that $|R'_1| = |R_1|$.   Then she replaces $R_1$ with $R_1'$.  For the second parent $R_2$, we compute $R_2'(\underline{A_2}, B, C, \share{T_2}) = R_2 \Join \pi_B(R_4)$ and output the results to Bob, who then replaces $R_2$ with $R_2'$ while renaming $B$ into $B^{(2)}$.  Afterwards, we add $B=B^{(2)}$ to $\Theta$ and remove $R_4$.  This completes the first iteration, with the updated schema shown in Figure~\ref{fig:it1}.  The reason we need to add $B=B^{(2)}$ as an additional equality constraint, which will be enforced after the join, is that the two binary joins $R_1 \Join R_4$ and $R_2 \Join \pi_B(R_4)$ have been performed separately, which does not check if the two $B$ attributes are equal. 

\begin{figure}
    \centering
    \begin{tikzpicture}[sibling distance=5em, level distance=3em,
      every node/.style = {shape=rectangle, draw, align=center}]]
    \node [color=blue] (root) at(0, 0) {$^{P_0\ }R_0(A_1, A_2, A_3)$};
    \node [color=blue] (left) at(-3, -1) {$^{P_0\ }R_1(\underline{A_1}, B, D, T_1)$};
    \node [color=red] (mid) at(0, -1) {$^{P_1\ }R_2(\underline{A_2}, B^{(2)}, C, \share{T_2})$};
    \node [color=red] (right) at(2.6, -1) {$^{P_1\ }R_3(\underline{A_3}, C)$};
    \node [color=blue] (rightleaf) at(1, -2) {$^{P_0\ }R_5(\underline{C}, E)$};
    \draw[->] (root) -- (left);
    \draw[->] (root) -- (mid);
    \draw[->] (root) -- (right);
    \draw[->] (mid) -- (rightleaf);
    \draw[->] (right) -- (rightleaf);
    \end{tikzpicture}
    \caption{Example~\ref{ex:2} after the first iteration.}
    \label{fig:it1}
\end{figure}
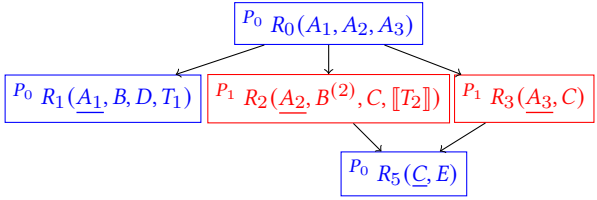

\begin{figure}
    \centering
    \begin{tikzpicture}[sibling distance=5em, level distance=3em,
      every node/.style = {shape=rectangle, draw, align=center}]]
    \node [color=blue] (root) at(0, 0) {$^{P_0\ }R_0(A_1, A_2, A_3)$};
    \node [color=blue] (left) at(-2.5, -1) {$^{P_0\ }R_1(\underline{A_1}, B, D, T_1)$};
    \node [color=red] (mid) at(0, -2) {$^{P_1\ }R_2(\underline{A_2}, B^{(2)}, C, \share{E}, \share{T_2})$};
    \node [color=red] (right) at(2.5, -1) {$^{P_1\ }R_3(\underline{A_3}, C^{(2)}, \share{T_3})$};
    \draw[->] (root) -- (left);
    \draw[->] (root) -- (mid);
    \draw[->] (root) -- (right);
    \end{tikzpicture}
    \caption{Example~\ref{ex:2} after the second iteration.}
    \label{fig:it2}
\end{figure}
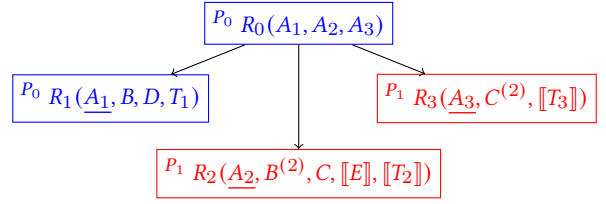

Suppose we pick $R_5$ in the second iteration.  $R_5$ also has two parents $R_2$ and $R_3$. We first compute $R_2'(\underline{A_2}, B^{(2)}, C, \share{E}, \share{T_2}) = R_2 \Join R_5$ and output the results to Bob.  Note that the indicators $T_2$ in $R_2'$ may be different from the original $T_2$ in $R_2$ in case not all tuples in $R_2$ can join with $R_5$, but we still have $|R_2'| = |R_2|$.  Bob then replaces $R_2$ with $R_2'$.  For the second parent $R_3$, we compute $R_3'(\underline{A_3}, C, \share{T_3}) = R_3 \Join \pi_C(R_5)$ and output the results to Bob, who then replaces $R_3$ with $R_3'$ while renaming $C$ into $C^{(2)}$.  Afterwards, we add $C=C^{(2)}$ to $\Theta$ and remove $R_5$.  The updated schema is shown in Figure~\ref{fig:it2}.

Now the join has become a tree join, and the remaining iterations are straightforward.  After the final iteration, the only table would have the form $J=(A_1, A_2, A_3, B, D, \share{B^{(2)}},  \share{C}, \share{E}, \share{C^{(2)}}, \share{T})$ held by Alice.  We use a garbled circuit to check $B=B^{(2)}, C=C^{(2)}$ in shared form, and set $\share{T} \leftarrow \share{T \cdot \Ind(B=B^{(2)}) \cdot \Ind(C=C^{(2)})}$, which yields the final join results. \qed
\end{example}
 
\paragraph{Cost analysis}
Given a database $\mathbf{R} = ( R_0(\mathbf{x}_0), R_1(\mathbf{x}_1), \dots, R_k(\mathbf{x}_k))$ conforming to an FK acyclic schema, our protocol computes a binary join $R\Join S$ for each parent-child relationship in the PK-FK DAG, whose cost is $O(|R|\log|R| +|S|\log|S|) $.  Note that $|R|$ does not change when it is replaced by the join results $R\Join S$.  Charging the $O(|R|\log |R|)$ term to $R$ and while the $O(|S|\log |S|)$ term to $S$, we conclude that the total cost for computing the whole join is $O\left(\sum_{i=0}^k d_i |R_i| \log|R_i|\right)$, where $d_i$ is the degree (sum of in-degree and out-degree) of $R_i$ in the PK-FK DAG.
 
\subsection{Supplementary Protocols}
Various relational operators can be applied to the join results before feeding them to the machine learning algorithm.  We discuss a few below.  We assume that Alice holds the join results.

\paragraph{Selection}
The selection operator selects a subset of rows using a predicate. As mentioned earlier, if the predicate involves attributes from one table, the party that owns the table may preprocess it before the join, possibly by adding dummy tuples if the selectivity of the predicate is sensitive information.  If a predicate involves attributes from different tables, we have to process it after the join.  We can process such a predicate in the same way we handled in equality constraints $\Theta$ above, by evaluating a garbled circuit for each row, and then multiplying its output to the indicator in shared form.  

\paragraph{Projection}
Note that there are two semantics of projection.  If duplicates are not removed, we can simply drop the attributes that are projected out.  If duplicates should be removed (as defined in relational algebra), this becomes a special group-by operation, as discussed next.

\paragraph{Group-by}
Suppose we would like to do a group-by on a set of attributes $\mathcal{A}$ followed by an aggregation on $B$. Let $\oplus$ be the aggregation operator, which can be any associative operation. We first sort the join results by $\mathcal{A}$.  If all attributes in $\mathcal{A}$ are in plaintext, Alice can do the sorting in plaintext, and then use OEP to permute $\share{B}$ and $\share{T}$ accordingly.  Otherwise, we use the sorting network to sort $(\share{\mathcal{A}}, \share{B}, \share{T})$ by $(T,\mathcal{A})$ such that all dummy tuples are placed after real tuples, and all real tuples that have the same values on $\mathcal{A}$ are consecutive.  

Suppose the join results are $\{(\share{a_i}, \share{b_i}, \share{t_i})\}_{i=1}^N$ after the sorting. 
The parties build a garbled circuit with $N-1$ \textit{merge gates}. The $i$-th merge gate first computes $z_i=t_{i+1}\cdot\Ind(a_i=a_{i+1})$ which indicates whether we should aggregate the values on $b_i$ and $b_{i+1}$ ($z_i=1$) or not ($z_i=0$). Then the gate updates $t_i\gets t_i(1-z_i)$ and
\begin{equation}\label{eq:agg}
    b_{i+1}\gets (b_i\oplus b_{i+1})z_i + b_{i+1}(1-z_i),
\end{equation} 
namely, when the aggregate should be performed, we set the $i$-th tuple to dummy and ``add'' $b_i$ to $b_{i+1}$.  Note that in \eqref{eq:agg}, the input $b_i$ is one of the outputs from the $(i-1)$-th merge gate when $i>1$, thus the depth of the circuit is $O(N)$.  Nevertheless, it can still be evaluated in constant number of rounds using Yao's garbled circuit.

A (distinct) projection operator corresponds to the case when $B=\emptyset$. In this case, we can simply ignore \eqref{eq:agg} in each merge gate.  The protocol would just set all tuples to dummy except  for one (the last one) tuple for each unique $a_i$. 

\subsection{The Two-Server Model}
\label{sec:twoserver}
In principle, our protocol can be generalized to more than two parties by just replacing the secret-sharing scheme and the atomic operations (addition and multiplication) to their multi-party counterparts. However, the cost increases significantly as more parties are added.  In practice, an alternative security model, known as the \textit{two-server model}, is more often used.  In the two-server model, we assume there are two semi-honest, non-colluding servers $P_0$ and $P_1$, and any number of parties, each holding a subset of the tables.  Each party first share the data to the two servers, i.e., for each tuple $t$, s/he sends $\share{t}_0$ to $P_0$ and sends $\share{t}_1$ to $P_1$.  Note that the two servers do not learn anything unless they collude.

Our protocol can be made to work in the two-server model, by using a modified version of the binary join protocol.  Recall that the protocol in Section~\ref{sec:binary} relies on the join attribute $B$ being in plaintext.  In the two-server model, all attributes are shared.  Specifically, consider the binary join $J=R(\share{\underline{A}}, \share{B})\Join S(\share{\underline{B}}, \share{C})$, where $S.B$ is an FK referencing $R.B$.  Let $|R|=M$ and $|S|=N$.  Since $B$ is shared, we can no longer 
use PSI.  Instead, we present a sorting based protocol as follows.
\begin{enumerate}
    \item \textbf{Extend:} We build a new table $J(\share{A},\share{B},\share{C},\share{U})$ with $M+N$ tuples. The first $M$ tuples are extended tuples from $R$, where we set $C=U=0$. The last $N$ tuples are extended tuples from $S$, where we set $A=0$, $U=1$.
    \item \textbf{Sort:} We use a sorting network to sort $J$ lexicographically by $(B,U)$ in descending order. After the sorting, tuples of $J$ with same value on $B$ are grouped together. In each group, the first tuple must be from $S$ if it exists.
    \item \textbf{Set indicators:} Let the $i$-th tuple of $J$ be $(a_i,b_i,c_i,u_i)$ after sorting. We will add an indicator attribute $T=\{t_i\}_{i=1}^{M+N}$ to $J$ while updating the $C$ attribute.  Set $t_1 = u_1$.  For $i=1,2,\cdots,M+N-1$, we update $c_{i+1} \leftarrow c_i$ and $t_{i+1} \gets t_i$ if $b_{i+1}=b_i$, and update $t_{i+1}\gets u_{i+1}$ if $b_{i+1}\neq b_i$.
    \item \textbf{Compact:} Finally, we sort $J$ again by $U$ in ascending order and the drop the $U$ column.  We only keep the first $M$ tuples. Then $J$ is the required join result.
\end{enumerate}

The cost of this protocol will increase from $O(M\log M + N \log N)$ to $O(M\log^2 M + N \log^2 N)$ due to sorting. Also, for simplicity we did not consider existing indicators in $R$ and $S$, which can be handled using similar ideas as in Section~\ref{sec:binary}.

\section{Secure Machine Learning}
\label{sec:ML}

\subsection{Join Result Purification}
Before feeding the join results to SecureML \cite{secureml}, we still need to remove the dummy tuples, i.e., rows where $T=0$.  For level 3 privacy protection, we can use oblivious compaction to move all real tuples to the front and only feed the first $|J|$ tuples to SecureML.  For level 4 privacy, however, this does not work since we are not allowed to reveal $|J|$. 

Below we design a purification circuit, which replaces all dummy tuples with duplicated real tuples. More precisely, in the ``purified'' $J$, every real tuple appears either $\left\lfloor{ |R_0| \over |J|} \right\rfloor$ or $\left\lfloor{ |R_0| \over |J|} \right\rfloor + 1$ times.  Then we do a random shuffle of the purified $J$ and feed it to SecureML to run minibatch SGD.

\begin{figure}[htbp]
    \centering
        \begin{tikzpicture}[sibling distance=5em, level distance=5em,
      every node/.style = {shape=rectangle, draw, align=center}]]
    \node  (x0) at(0, 0) {$X$};
    \node  (x1) at(1, 0) {$Y$};
    \node  (x2) at(2, 0) {$D$};
    \node  (x3) at(3, 0) {$D$};
    \node  (x4) at(4, 0) {$D$};
    \node  (x5) at(5, 0) {$D$};
    \node  (x6) at(6, 0) {$D$};
    \node  (x7) at(7, 0) {$D$};
    
    \node  (y0) at(0, -1) {$X$};
    \node  (y1) at(1, -1) {$Y$};
    \node  (y2) at(2, -1) {$D$};
    \node  (y3) at(3, -1) {$D$};
    \node  (y4) at(4, -1) {$X$};
    \node  (y5) at(5, -1) {$Y$};
    \node  (y6) at(6, -1) {$D$};
    \node  (y7) at(7, -1) {$D$};
    
    \node  (z0) at(0, -2) {$X$};
    \node  (z1) at(1, -2) {$Y$};
    \node  (z2) at(2, -2) {$X$};
    \node  (z3) at(3, -2) {$Y$};
    \node  (z4) at(4, -2) {$X$};
    \node  (z5) at(5, -2) {$Y$};
    \node  (z6) at(6, -2) {$X$};
    \node  (z7) at(7, -2) {$Y$};
    
    \node  (w0) at(0, -3) {$X$};
    \node  (w1) at(1, -3) {$Y$};
    \node  (w2) at(2, -3) {$X$};
    \node  (w3) at(3, -3) {$Y$};
    \node  (w4) at(4, -3) {$X$};
    \node  (w5) at(5, -3) {$Y$};
    \node  (w6) at(6, -3) {$X$};
    \node  (w7) at(7, -3) {$Y$};
    
    \draw[->] (x0) -- (y0);
    \draw[->] (x0) -- (y4);
    \draw[->] (x1) -- (y1);
    \draw[->] (x1) -- (y5);
    \draw[->] (x2) -- (y2);
    \draw[->] (x2) -- (y6);
    \draw[->] (x3) -- (y3);
    \draw[->] (x3) -- (y7);
    \draw[->] (x4) -- (y4);
    \draw[->] (x5) -- (y5);
    \draw[->] (x6) -- (y6);
    \draw[->] (x7) -- (y7);
    
    \draw[->] (y0) -- (z0);
    \draw[->] (y0) -- (z2);
    \draw[->] (y1) -- (z1);
    \draw[->] (y1) -- (z3);
    \draw[->] (y2) -- (z2);
    \draw[->] (y2) -- (z4);
    \draw[->] (y3) -- (z3);
    \draw[->] (y3) -- (z5);
    \draw[->] (y4) -- (z4);
    \draw[->] (y4) -- (z6);
    \draw[->] (y5) -- (z5);
    \draw[->] (y5) -- (z7);
    \draw[->] (y6) -- (z6);
    \draw[->] (y7) -- (z7);
    
    \draw[->] (z0) -- (w0);
    \draw[->] (z0) -- (w1);
    \draw[->] (z1) -- (w1);
    \draw[->] (z1) -- (w2);
    \draw[->] (z2) -- (w2);
    \draw[->] (z2) -- (w3);
    \draw[->] (z3) -- (w3);
    \draw[->] (z3) -- (w4);
    \draw[->] (z4) -- (w4);
    \draw[->] (z4) -- (w5);
    \draw[->] (z5) -- (w5);
    \draw[->] (z5) -- (w6);
    \draw[->] (z6) -- (w6);
    \draw[->] (z6) -- (w7);
    \draw[->] (z7) -- (w7);

    \end{tikzpicture}
    \caption{A duplication circuit}
    \label{duplicationnetwork}
\end{figure}
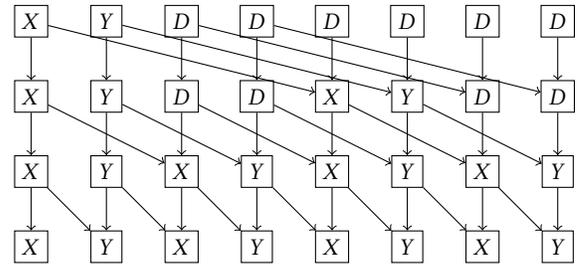

We first use oblivious compaction to move all real tuples to the front of $J$.  Then we use a duplication circuit to replace all dummy tuples with real ones, as shown in Figure~\ref{duplicationnetwork} where $D$ represents a dummy tuple.  Let $N=|R_0|$ be the join results including dummy tuples. This circuit contains $\ell = \lfloor\log N\rfloor$ levels. 
Let $x_j^{(i)}$ be the $j$-th tuple after the $i$-th level of the circuit for $i=0,1,\dots,\ell$ and $j=1,2,\dots,|R_0|$, where $\{x_j^{(0)}\}$ are the inputs of the circuit and $\{x_j^{(\ell)}\}$ are the outputs of the circuit. Then for $i=1,\dots,\ell$, the gates in the $i$-th level are described as
\[
x_j^{(i)}=
\begin{cases}
x_j^{(i-1)}&\text{ if }j\leq 2^{\ell-i}\text{ or }x_j^{(i-1)}\text{ is real,}\\
x_{j-2^{\ell-i}}^{(i-1)}&\text{ otherwise.} 
\end{cases}
\]

However, the duplication circuit in Figure~\ref{duplicationnetwork} only works when $|J|$ is a power of 2.  When $|J|$ is not a power of $2$, some tuples may be duplicated much more than the others.  For example, suppose $N=16$ and $|J|=3$.  After the compaction we have the sequence $(X, Y, Z, D, D, D, D, D, D, D, D, D, D, D, D, D)$, where $D$ represents a dummy tuple. After the duplication network, the result would be $(X, Y, Z, X, X, Y, Z, X, X, Y, Z, X, X, Y, Z, X)$, in which $X$ appears 8 times while $Y, Z$ appear 4 times each, which will cause biases in the training. 

\begin{figure}[htbp]
    \centering
        \begin{tikzpicture}[sibling distance=5em, level distance=5em,
      every node/.style = {shape=rectangle, draw, align=center}]]
    \node  (x0) at(0, 0) {$X$};
    \node  (x1) at(1, 0) {$Y$};
    \node  (x2) at(2, 0) {$Z$};
    \node  (x3) at(3, 0) {$U$};
    \node  (x4) at(4, 0) {$V$};
    \node  (x5) at(5, 0) {$D$};
    \node  (x6) at(6, 0) {$D$};
    \node  (x7) at(7, 0) {$D$};
    
    \node  (y0) at(0, -1) {$X$};
    \node  (y1) at(1, -1) {$Y$};
    \node  (y2) at(2, -1) {$Z$};
    \node  (y3) at(3, -1) {$U$};
    \node  (y4) at(4, -1) {$V$};
    \node  (y5) at(5, -1) {$Y$};
    \node  (y6) at(6, -1) {$Z$};
    \node  (y7) at(7, -1) {$U$};
    
    \draw[->] (x0) -- (y0);
    \draw[->] (x0) -- (y4);
    \draw[->] (x1) -- (y1);
    \draw[->] (x1) -- (y5);
    \draw[->] (x2) -- (y2);
    \draw[->] (x2) -- (y6);
    \draw[->] (x3) -- (y3);
    \draw[->] (x3) -- (y7);
    \draw[->] (x4) -- (y4);
    \draw[->] (x5) -- (y5);
    \draw[->] (x6) -- (y6);
    \draw[->] (x7) -- (y7);
    
    \end{tikzpicture}
    \caption{half copy network}
    \label{halfcopynetwork}
\end{figure}
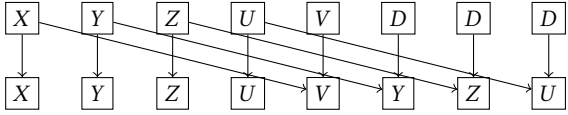

To address the issue, we only run the duplication circuit for $\lfloor \log(N/|J|) \rfloor$ levels. As $|J|$ is sensitive, what we do more precisely is that for each $i> \lfloor \log(N/|J|) \rfloor$, we set $x_j^{(i)}=x_j^{(i-1)}$ for all $j$.  Following the earlier example, this would result in the sequence $(X, Y, Z, D, X, Y, Z, D, X, Y, Z, D, X, Y, Z, D)$.  After this, we compact the sequence again, putting the real tuples at the front.  Note that we now must have at least $N/2$ real tuples.  Finally, we use a half-copy network shown in Figure~\ref{halfcopynetwork}, in which $x_{j}$ is replaced with $x_{j-N/2}$ if $x_{j}$ is dummy, for $j>N/2$.  This makes sure that the numbers of copies of each real tuple differ by at most one, which unavoidably happens when $N$ is not a multiple of $|J|$.

\subsection{Privatizing Trained Model }
\label{sec:DP}
The ML model trained by SecureML is stored in shared form.  So if we adopt join result purification, no information at all is revealed so far, which also means that neither party has learned anything.  In the MPC model, the trained model can be revealed, since it is the output of the functionality.  Doing so corresponds to level 4 privacy protection.

However, the trained model itself inevitably contains information about the data.  To achieve level 5 privacy protection, we use differential privacy (DP) to inject noise into the model.  Towards this goal, we adopt the popular \textit{gradient perturbation} approach \cite{Abadi_2016} to add Gaussian noise to the gradient of each minibatch.  More precisely, for each minibatch $B$, we update the coefficients as 
\begin{equation}
\label{eq:DP}
\mathbf{w}_{i+1} \gets \mathbf{w}_i - \frac{\eta}{|B|} \left( \sum_{t \in B}   \nabla L(t, \mathbf{w}_i)  + \tau \sigma C \cdot \mathcal{N}(0, \mathbf{I}) \right).
\end{equation}
Here, $\mathcal{N}(0, \mathbf{I})$ represents a vector of random variables, each of which is drawn from the Gaussian distribution with mean $0$ and variance $1$.  The parameter $C$ is the $\ell_2$ clipping threshold of the gradient and $\sigma = {|B| \over |J|}\sqrt{T\log(1/\delta)\log(T/\delta)}/\varepsilon$, where $T$ is the number of minibatches, and $\delta, \varepsilon$ are the DP parameters.  Compared with \citet{Abadi_2016}, we need an extra multiplier $\tau$, which is the maximum number of join results any tuple in the primary private table can produce.  For example, in the TPC-H schema, if $\mathtt{customer}$ is the primary private table, $\tau$ is the maximum number of lineitems any customer has purchased.  Recall that under the DP policy over relational data \cite{kotsogiannis2019privatesql}, all the join results associated with one tuple in the primary private table may be deleted to obtain a neighboring instance, so we need to scale up the noise level by $\tau$ in order to protect the privacy of tuples in the primary private table.  We assume that the training hyper-parameters $|B|, T, \eta$, as well as the DP parameters $\varepsilon, \delta$, and $\tau$ are public knowledge.

The remaining technical issue is how to implement \eqref{eq:DP} in the MPC model.  First, note that $|J|$ is secret, so we first compute $\share{\sigma}$ in shared form.  This requires an MPC division \cite{EMPtoolkit, ABYframework}, but we only need to do it once.  Next, SecureML can compute the minibatch gradient sum $\share{\sum_{t\in B}\nabla L(t, \mathbf{w}_i)}$ in shared form, so it only remains to show how generate noise.  To do so, we use a technique from \cite{NEURIPS2018_7221e5c8}. Alice and Bob each generates a uniform number in $\mathbb{Z}_n$.  Note that these two values form an arithmetic share of a uniformly random variable.  Then we convert it to a Gaussian random variable from $\mathcal{N}(0,1)$ inside MPC.  Finally, we multiply it with $\share{\tau \sigma C}$ and add to $\share{\sum_{t\in B}\nabla L(t, \mathbf{w}_i)}$.

\section{Experiments}

\subsection{Experimental Setup}
We implemented our prototype system under the ABY \cite{ABYframework}, which provides efficient conversions between different secret-sharing schemes, together with common cryptographic operations in the secure two-party setting.  The system consists of three stages.  The first stage, called  SFKJ (standing for Secure Foreign-Key Join), computes the join results, together with selection predicates if there are any, in secret-shared form.  The second stage is join result purification, which is needed for level 4 privacy. The third stage is SecureML \cite{secureml}, whose code currently supports linear regression and logistic regression (neural networks can be also supported  \cite{secureml}, but it is not available in the released code).  We augment the SecureML code with an option of DP noise injection to reach level 5 privacy, following the description in Section~\ref{sec:DP}.  

We ran all experiments on two VMs (Standard\_D8s\_v3, 8vCPU, 32GB RAM) on Azure under two scenarios: In the WAN setting, one VM is located in Eastern US and the other in East Asia. The average network latency is around 210ms between two VMs and the network bandwidth is around 100Mb/s.  In the LAN setting, both VMs are in the same region, with a network delay around 0.1ms and bandwidth around 1Gb/s.
The computational security parameter is set to $\kappa = 128$ and the statistical security parameter is set to $\sigma = 40$. The bit length of all attributes is set to $l = 32$ during the SFKJ and purification stage, while we convert data (in shared form) to $l = 64$ for achieving a better accuracy before feeding it to SecureML. 

\subsection{SFKJ}
We used the TPC-H dataset and a real-world dataset MovieLens, which stores movie recommendations, to test the performance of SFKJ.  We benchmark SFKJ with the following three options:

\begin{itemize}
\item Plain text: In this baseline approach, one party sends all her data to the other party, who computes the join on plain text. 
\item Garbled circuit: The most widely used generic MPC protocol in the two-party setting is Yao's garbled circuit \cite{yao1982protocols}.  To implement a multi-way join using a garbled circuit, one baseline approach is to use a circuit that compares all combinations of the tuples, one from each relation, and checks if they satisfy all join and selection conditions. This is the approach taken by SMCQL \cite{bater2017smcql}.
\item Optimized garbled circuit: For FK acyclic joins, we observe that the join size is bounded by the size of the fact table.  Thus, an improved circuit design is to perform the multi-way join in a pairwise fashion.  After each two-way join, we compact (using the compaction circuit) the intermediate join results to the size of the fact table.

\end{itemize}

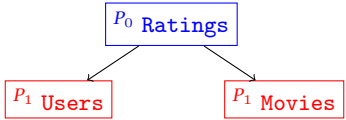
\begin{figure}[htbp]
    \centering
        \begin{tikzpicture}[sibling distance=5em, level distance=3em,
          every node/.style = {shape=rectangle, draw, align=center}]]
        \node [color=blue] (root) at(0, 0) {$^{P_0\ }\mathtt{Ratings}$};
        \node [color=red] (left) at(-1.5, -1) {$^{P_1\ }\mathtt{Users}$};
        \node [color=red] (right) at(1.5, -1) {$^{P_1\ }\mathtt{Movies}$};
        \draw[->] (root) -- (left);
        \draw[->] (root) -- (right);
        \end{tikzpicture}
    \caption{Foreign-key graph of MovieLens}
    \label{fig:movielens}
\end{figure}

\paragraph{MovieLens} 
The dataset contains 3 tables, $\mathtt{Users}$ (userID, age, gender and other personal information), $\mathtt{Movies}$ (movieID, movie year and its type), and $\mathtt{Ratings}$ (userID, movieID, score, timestamp) that users gave to movies. $\mathtt{Ratings.userID}$ and $\mathtt{Ratings.movieID}$ are the foreign keys referencing $\mathtt{Users}$ and $\mathtt{Movies}$ respectively. The schema is shown in Figure \ref{fig:movielens}, where Alice ($P_0$) holds the fact table $\mathtt{Ratings}$, and Bob ($P_1$) holds $\mathtt{Users}$ and $\mathtt{Movies}$.  

\begin{figure}[htbp]
\subfigure[Communication cost (MB)] 
{
	\begin{minipage}{4cm}
	\centering 
	\includegraphics[scale=0.4]{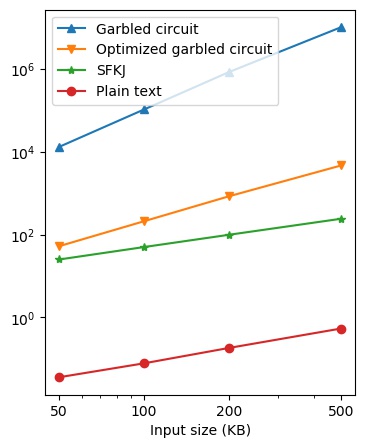}
	\end{minipage}
}
\subfigure[Running time (s)] 
{
	\begin{minipage}{4cm}
	\centering
	\includegraphics[scale=0.4]{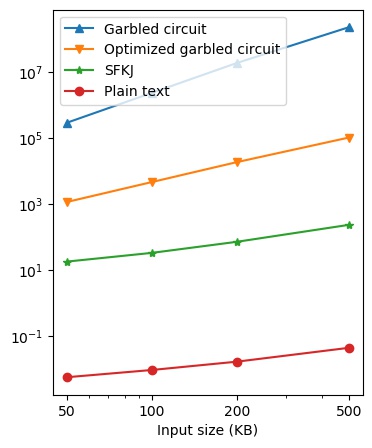}
	\end{minipage}
}
\caption{Experiment results on the MovieLens dataset}
\label{fig:exp_movielens}
\end{figure}

Figure \ref{fig:exp_movielens} reports the communication costs and running time of the 4 methods as we vary the data size in the LAN setting.  Both SFKJ and the plain text method have linear growth in terms of communication cost and CPU time.  On the other hand, the garbled circuit and optimized garbled circuit grow at a faster rate.  Note that both axes are drawn in log scale, so a higher aspect ratio indicates a polynomially faster rate.  Indeed, the optimized garbled circuit has a quadratic growth rate (due to computing the Cartesian product of every two-way join), while the garbled circuit has a cubic growth rate (it computes the Cartesian product of three tables). In fact, we could not run the garbled circuit on the largest dataset, which would take more than 3 months.  The reported running times and communication costs are extrapolated from the result on smaller datasets.  This is actually very accurate, since these costs are precisely proportional to the circuit size, which we can compute. 

We omit the results in the WAN setting.  In this case the running time is dominated by the communication, which is roughly the total communication cost divided by the bandwidth.  The network delay has a negligible effect, since SFKJ has a constant number (only depending on the number of tables in the join and independent of data size) of communication rounds.

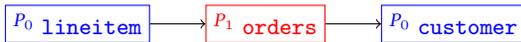
\begin{figure}[htbp]
    \centering
        \begin{tikzpicture}[sibling distance=5em, level distance=3em,
          every node/.style = {shape=rectangle, draw, align=center}]]
        \node [color=blue] (root) at(0, 0) {$^{P_0\ }\mathtt{lineitem}$};
        \node [color=red] (middle) at(2.5, 0) {$^{P_1\ }\mathtt{orders}$};
        \node [color=blue] (leaf) at(5, 0) {$^{P_0\ }\mathtt{customer}$};
        \draw[->] (root) -- (middle);
        \draw[->] (middle) -- (leaf);
        \end{tikzpicture}
    \caption{Foreign-key graph of TPC-H Query 3}
    \label{fig:tpchq3}
\end{figure}

\paragraph{TPC-H} We tried two queries on the TPC-H dataset.  The first query, Q3, involves a 3-table line join as shown in Figure \ref{fig:tpchq3}. We let Alice ($P_0$) hold $\mathtt{lineitem}$ and $\mathtt{customer}$, and let Bob ($P_1$) hold $\mathtt{orders}$.

\begin{figure}[htbp]
    \centering
    \begin{tikzpicture}[sibling distance=5em, level distance=3em,
      every node/.style = {shape=rectangle, draw, align=center}]]
    \node [color=blue] (root) at(0, 0) {$^{P_0\ }\mathtt{lineitem}$};
    \node [color=red] (left) at(2, 0.9) {$^{P_1\ }\mathtt{supplier}$};
    \node [color=red] (rightup) at(1, -0.9) {$^{P_1\ }\mathtt{orders}$};
    \node [color=red] (rightdown) at(3.5, -0.9) {$^{P_1\ }\mathtt{customer}$};
    \node [color=blue] (leaf) at(4, 0) {$^{P_0\ }\mathtt{nation}$};
    \node [color=blue] (bottom) at(6, 0) {$^{P_0\ }\mathtt{region}$};
    \draw[->] (root) -- (left);
    \draw[->] (root) -- (rightup);
    \draw[->] (rightup) -- (rightdown);
    \draw[->] (rightdown) -- (leaf);
    \draw[->] (left) -- (leaf);
    \draw[->] (leaf) -- (bottom);
    \end{tikzpicture}
    \caption{Foreign-key graph of TPC-H Query 5}
    \label{fig:tpchq5}
\end{figure}
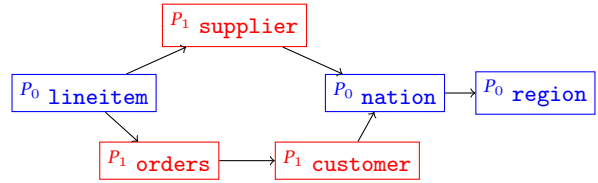

The second query is Q5, which features a more complicated join structure, as shown in Figure~\ref{fig:tpchq5}. We let Alice ($P_0$) hold $\mathtt{lineitem}$, $\mathtt{nation}$, and $\mathtt{region}$, and let Bob ($P_1$) hold $\mathtt{supplier}$, $\mathtt{orders}$, and $\mathtt{customer}$. We use this query to illustrate possible optimizations: Alice and Bob can first locally compute $\mathtt{NR} = \mathtt{nation} \Join \mathtt{region}$ and $\mathtt{OC} = \mathtt{orders} \Join \mathtt{customer}$ in plain text.  Then we compute $\mathtt{supplier'} = \mathtt{supplier} \Join \mathtt{NR}$ and $\mathtt{OC'} = \mathtt{OC} \Join \mathtt{NR}$ using our secure join protocol.  This query is a non-tree join, so we also need to add a constraint $\mathtt{supplier'.nationkey} = \mathtt{OC'.nationkey}$ into $\Theta$, which is enforced by an extra garbled circuit in the end. Finally, we perform a basic tree join on three tables $\mathtt{lineitem} \Join \mathtt{supplier'} \Join \mathtt{OC'}$.

\begin{figure}[htbp]
\subfigure[Communication Cost (MB)] 
{
	\begin{minipage}{4cm}
	\centering 
	\includegraphics[scale=0.4]{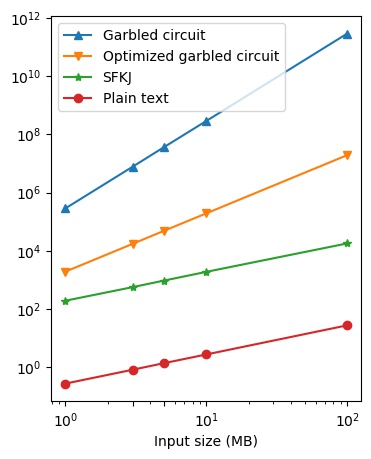}
	\end{minipage}
}
\subfigure[Running time (s)] 
{
	\begin{minipage}{4cm}
	\centering
	\includegraphics[scale=0.4]{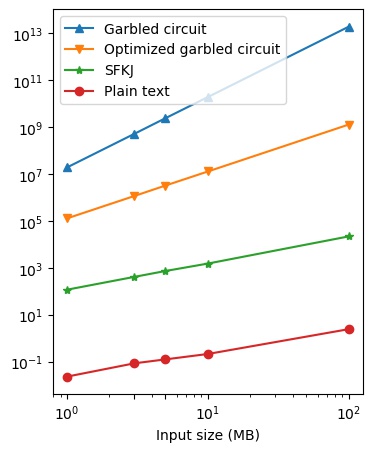}
	\end{minipage}
}
\caption{Experiment results on TPC-H Query 3}
\label{fig:exp_Q3}
\end{figure}

\begin{figure}[htbp]
\subfigure[Communication Cost (MB)] 
{
	\begin{minipage}{4cm}
	\centering 
	\includegraphics[scale=0.42]{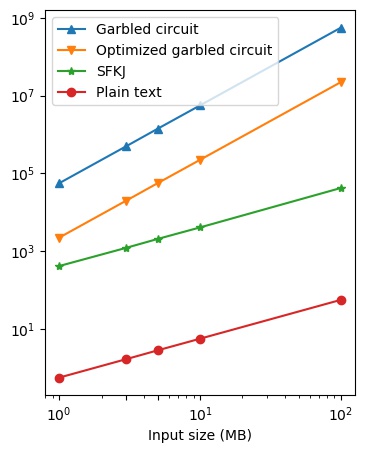}
	\end{minipage}
}
\subfigure[Running time (s)] 
{
	\begin{minipage}{4cm}
	\centering
	\includegraphics[scale=0.42]{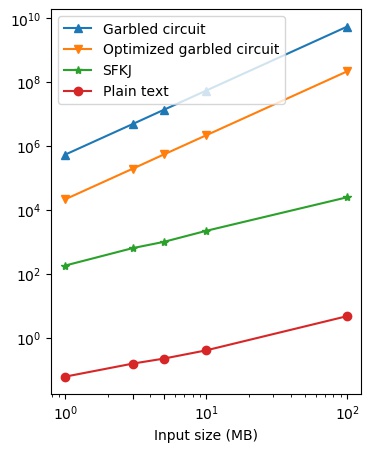}
	\end{minipage}
}
\caption{Experiment results on TPC-H Query 5}
\label{fig:exp_Q5}
\end{figure}

The experimental results in the LAN setting are shown in Figure \ref{fig:exp_Q3} and \ref{fig:exp_Q5}.  They resemble the results in Figure \ref{fig:exp_movielens}, except that the gap between the garbled circuits and SFKJ is even larger, due to the larger scale of the datasets.  For example, on the 100 MB dataset with Query 5, SFKJ takes 6.8 hours while the optimized garbled circuit is estimated to take 7 years.

\subsection{Join Result Purification}
To test the efficacy of join result purification for machine learning, we used the $10^4$ handwritten digits from the MNIST dataset \cite{mnist} under the following setting with predicates.  We first added (randomly generated) IDs to the MNIST dataset as PKs, and then split the feature columns (there are 784 features) and the label column into two tables, held by Alice and Bob, respectively.  Suppose Alice and Bob each has an extra column $\mathtt{flag}$ that indicates whether s/he would like to include a particular record for the training, and the training is done on the PK-PK join results with the predicate $\mathtt{flag=True}$. 
For the experiments, we let Alice and Bob randomly set $\mathtt{flag}$ to $\mathtt{True}$ with probability $1/3$, thus about $1/9$ of the entire dataset were used for training. Recall that a record can appear in the join result only when both parties agree to use it; otherwise, it will appear in the join results as a dummy tuple.


We use the MNIST dataset to perform a binary classification (whether the hand-written number is 0 or not) using logistic regression. 
The join takes 306 seconds (in the LAN setting) and 98 MB of communication.  Afterwards, we feed the join results to SecureML, first including all dummy tuples, and then with the purified results. The purification takes 133 seconds in the LAN setting, and 252 seconds in the WAN setting, which is basically dominated by the 2170 MB of communication.

\begin{figure}[htbp]
     \centering
     \includegraphics[scale=0.6]{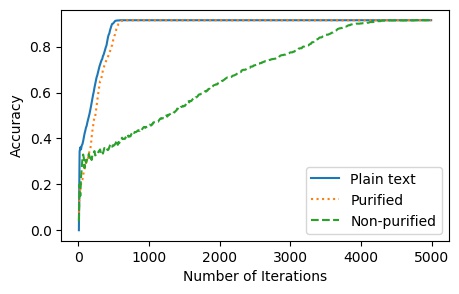}
     \caption{Training progress on MNIST}
     \label{fig:iter-acc-mnist}
 \end{figure}

Figure \ref{fig:iter-acc-mnist} shows the training progress on join results with dummy tuples as well as on purified results.  We also benchmark them with plaintext training.  We see that on purified join results, the training progress is almost as good as that on the plain text, which shows a high efficacy of the purification stage. 
Table \ref{tab:mnisttraincost} shows the time and communication cost of training phase.  Recall that SecureML uses multiplication triples, which can be generated in an offline stage.  Note that the offline cost still has to be borne by the two parties, just that they can choose to do it during non-peak hours.  From the results, we see that purification saved around 3500 iterations of training, which is definitely worthy of its cost.  In particular, since each iteration requires a few rounds of communication, 
the savings are more significant in the WAN setting due to the network delay.

\begin{table}[htbp]
\begin{tabular}{cccc}
\hline
 & Time (LAN) & Time (WAN) & Commun. \\
\hline
Purification & 133.52s & 252.13s & 2.17 GB\\
Offline (per iter.) & 4.02s & 45.2s & 392 MB\\
Online (per iter.) & 0.03s & 1.26s & 20.25 KB \\
Offline (3500 iter.) & 3.91h & 44h & 1.31 TB\\
Online (3500 iter.) & 105s & 1.23h & 69.21 MB \\
\hline
\end{tabular}
\caption{MNIST Training Cost}
\label{tab:mnisttraincost}
\end{table}

\subsection{The Full Pipeline}
\label{allphase}

\begin{figure}[htbp]
    \centering
        \begin{tikzpicture}[sibling distance=5em, level distance=3em,
          every node/.style = {shape=rectangle, draw, align=center}]]
        \node [color=blue] (root) at(0, 0) {$^{P_0\ }\mathtt{studentInfo}$};
        \node [color=red] (left) at(-1.5, -1) {$^{P_1\ }\mathtt{studentVle}$};
        \node [color=red] (right) at(1.5, -1) {$^{P_1\ }\mathtt{studentAssessment}$};
        \node [color=red] (down) at(1.5, -2) {$^{P_1\ }\mathtt{assessments}$};
        \draw[->] (root) -- (left);
        \draw[->] (root) -- (right);
        \draw[->] (right) -- (down);
        \end{tikzpicture}
    \caption{Foreign-key graph of OULAD}
    \label{fig:oulad}
\end{figure}
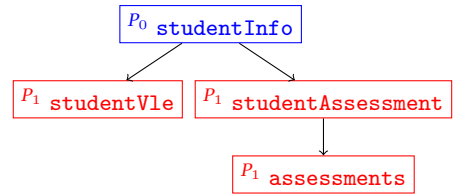

Finally, we test the join-purification-learning pipeline using the Open University Learning Analytics dataset (OULAD) \cite{oulad}, which contains data about courses, students, and their interactions in a virtual learning environment (Vle).  We used 4 tables in the dataset and partitioned them to Alice ($P_0$) and Bob ($P_1$) as shown in Figure~\ref{fig:oulad}.  Table $\mathtt{studentAssessment}$ contains students' exam grades and assignment grades; $\mathtt{studentVle}$ contains the number of student’s interactions with the materials; and $\mathtt{studentInfo}$ contains the number times the student has attempted this module and the final results (``distinction'', ``pass'', and ``fail'').  There are $45,000$ tuples in all the tables. Our task is to predict the students' final results based on their performance using linear regression. We set batch size as 128 and the training takes 3000 iterations. The time and communication costs of each stage are listed in Table \ref{tab:overallview}.

 \begin{table}[htbp]
\begin{tabular}{cccc}
\hline
Stage               & Time(LAN)    & Time(WAN)  & Communication \\ \hline
SFKJ        & 596.64s  & 705.47s & 587.50 MB \\ 
Purification        & 86.82s   & 103.31s & 149.22 MB\\
Generate DP noises   & 296.85s   & 837.69s & 32.07 MB\\
SecureML(offline)   & 114.20s  & 1216.79s & 1439.36 MB\\
SecureML(online)    & 1.161s    & 1272.81s & 7.97 MB \\ \hline

\end{tabular}
\caption{Costs of different stages on the OULAD dataset}
\label{tab:overallview}
\end{table}

We have also examined the accuracy loss due to MPC and DP, respectively.  When injecting DP noise, we set $\tau =1$ (the students are considered individuals whose privacy is to be protected by DP and each student has at most one record in one course), $C = 1$, $\delta=10^{-6}$, and tried different values for $\varepsilon$ (recall that lower $\varepsilon$ means higher privacy requirement).  The results are reported in Table~\ref{tab:accuracy}, where $\varepsilon=\infty$ corresponds to level 4 privacy protection, i.e., without DP noise. We see that for $\varepsilon<1$, the DP noise incurs some non-negligible loss to the accuracy.  Running the noise injection in MPC introduces some small additional losses\footnote{In fact, in certain cases, the model computed in the MPC has (slightly) better accuracy than on plaintext.  This is because in some cases, the rounding errors actually cancel the DP noise, resulting in a (slightly) better model.} due to rounding errors (we used 12 bits of precision for the decimal part).  Note that the rounding errors can be further reduced by using higher precision in the MPC, but the loss due to DP in inevitable, unless better DP training algorithms are invented.

\begin{table}[htbp]
\begin{tabular}{c|c|c}
\hline
$\varepsilon$ & MPC Acc. & Plain Acc. \\ \hline
0.1  & 52.79\% & 52.26\% \\
0.2  & 68.71\% & 71.58\% \\
0.5  & 80.85\% & 80.65\% \\
1 &  83.21\% & 84.93\% \\
2 &  85.19\% & 86.08\% \\
5 & 86.20\% & 86.40\%\\
10 & 86.36\% & 86.57\%\\ 
$\infty$  &  86.44\% & 86.81\% \\  
\hline
\end{tabular}
\caption{Learning Accuracy with DP}
\label{tab:accuracy}
\end{table}

\bibliographystyle{ACM-Reference-Format}
\bibliography{ref}

\end{document}